\pdfoutput=1 
\documentclass[%
pra,reprint
]{revtex4-1}

\usepackage{amsmath}
\usepackage{graphicx}

\begin{document}

\preprint{APS/123-QED}

\title{Microwave-free magnetometry with nitrogen-vacancy centers in diamond}

\author{Arne Wickenbrock}
\affiliation{Johannes Gutenberg-Universit{\"a}t  Mainz, 55128 Mainz, Germany}
\email{wickenbr@uni-mainz.de}
\author{Huijie Zheng}
\affiliation{Johannes Gutenberg-Universit{\"a}t  Mainz, 55128 Mainz, Germany}
\author{Lykourgos Bougas}
\affiliation{Helmholtz Institut Mainz, 55099 Mainz, Germany}
\author{Nathan Leefer}
\affiliation{Helmholtz Institut Mainz, 55099 Mainz, Germany}
\author{Samer Afach}
\affiliation{Johannes Gutenberg-Universit{\"a}t  Mainz, 55128 Mainz, Germany}
\author{Andrey Jarmola}
\affiliation{Department of Physics, University of California, Berkeley, CA 94720-7300, USA}
\author{Victor M. Acosta}
\affiliation{Department of Physics and Astronomy, University of New Mexico, Center for High Technology Materials, Albuquerque, NM, 87106, USA}
\author{Dmitry Budker}
\affiliation{Johannes Gutenberg-Universit{\"a}t  Mainz, 55128 Mainz, Germany}
\affiliation{Helmholtz Institut Mainz, 55099 Mainz, Germany}
\affiliation{Department of Physics, University of California, Berkeley, CA 94720-7300, USA}
\affiliation{Nuclear Science Division, Lawrence Berkeley National Laboratory, Berkeley, CA 94720, USA}



\date{\today}

\begin{abstract}
We use magnetic-field-dependent features in the photoluminescence of negatively charged nitrogen-vacancy centers to measure magnetic fields without the use of microwaves. In particular, we present a magnetometer based on the level anti-crossing in the triplet ground state at 102.4\,mT with a demonstrated noise floor of 6\,nT/$\sqrt{\text{Hz}}$, limited by the intensity noise of the laser and the performance of the background-field power supply. The technique presented here can be useful in applications where the sensor is placed closed to conductive materials, e.g. magnetic induction tomography or magnetic field mapping, and in remote-sensing applications since principally no electrical access is needed.
\end{abstract}

\maketitle

\begin{figure}
  \centering
  \includegraphics[width=\columnwidth]{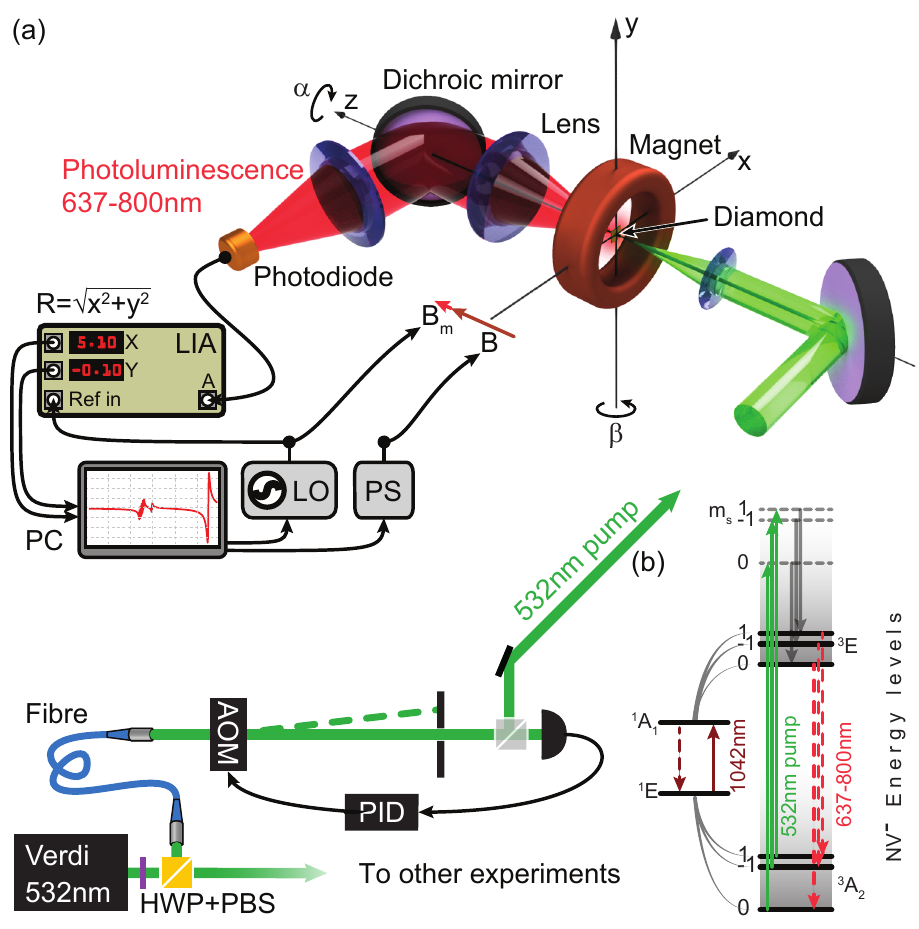}
\caption{(a) Schematic of the experimental setup. (b) NV-center energy level schematic.}\label{Figure1}
\end{figure}

The negatively-charged nitrogen-vacancy (NV) center in diamond has emerged as a unique nanoscale sensor full of interesting applications, and has been extensively researched in the past years resulting in numerous technological breakthroughs. It forms the basis for sensors to detect magnetic fields\cite{Rondin2014}, temperature\cite{SingleNVThermometry, TempBudker}, strain\cite{Strain}, rotation\cite{GyroLedbetter,RotationalSensing}, electric fields\cite{SingleNVElectricFieldSensing}, and quantum geometrical phases\cite{GeometricPhaseSensor}. In particular, the use of the optically detected magnetic resonance (ODMR) technique\cite{ODMR1} to probe the magnetically sensitive ground state of the NV center has proven to be a successful tool for sensitive magnetic field measurements, both with single and ensembles of NV centers\cite{Rondin2014,Wolf2015,SingleNVMagnetometry}. Realization of ODMR sensing protocols typically involves green pump light for optical polarization of the NV centers, the application of microwave (MW) fields for the manipulation of their spin state, and an optical readout step involving either detection of NV-photoluminescence (PL)\cite{Rondin2014} or absorption of 1042\,nm radiation resonant with the singlet transition\cite{Budker1,AcostaPRB2010,NVCavity1,Cavity2}. The relevant energy levels can be seen in Fig.\,\ref{Figure1}\,\cite{Rondin2014}. When the applied MW fields are resonant with the splitting of the Zeeman sub-levels of the NV-center, the transfer of spin populations results in an observable change in PL or absorption. 

\indent There are cases where the use of strong MW fields proves to be detrimental to the sensing protocol and therefore can prohibit the employment of an NV-based sensor. An example is the detection of magnetic fields generated by eddy currents in conductive materials in the context of magnetic induction tomography\cite{MIT2,MIT3a} (MIT), a research application currently undertaken in our laboratory\cite{MIT3}, where the presence of a conductive structure under examination will heavily affect the application of MW to the diamond. Another example is magnetic field mapping of conductive, magnetic structures\cite{Simpson2016}.\\
\indent There have been several demonstrations of MW-free, and all-optical, diamond-based magnetic sensors, initially implemented with single NV centers attached to scanning atomic force microscopes\,\cite{MaletinskyP2012,RondinAPL2012,Singlet2}, and more recently with ensembles of NV centers\,\cite{Simpson2016}. These MW-free magnetometric protocols have been realized by exploiting either the properties of the NV-centers' PL or their decoherence properties under the influence of external magnetic fields. So far, these protocols remain either qualitative, requiring complicated setups to achieve high spatial resolution, or lack high magnetic-field sensitivities and bandwidth. \\
\indent In this Letter, we demonstrate the principles of a sensitive MW-free magnetometer by exploiting the properties of the ground-state level anti-crossing (GSLAC) of the NV center in diamond. We note that the presented technique can be extended to other magnetically sensitive features in the NV-PL or absorption, as discussed later, as well as features associated with other spin defects in solid-state systems. 
In particular, for the NV center, a $\sim$102.4\,mT background magnetic field causes ground-state Zeeman-sublevel degeneracy and mixing (the anti-crossing), which is visible as a drop in NV-PL under optical pumping.
Any additional external magnetic field will perturb the anti-crossing condition and, thus, result in a PL change that can be monitored and used for sensitive detection of the perturbing magnetic field.\\ 
\indent Using this technique, we demonstrate a MW-free magnetometer with a 6\,nT/$\sqrt{\rm{Hz}}$ magnetic field sensitivity, a bandwidth exceeding 300\,kHz, and a projected 0.43\,nT/$\sqrt{\rm{Hz}}$ sensitivity limited by the photon shot noise of the PL detection\,\cite{AcostaPRL2013}.\\ 
A schematic of the experimental setup is shown in Fig.~\ref{Figure1}. We used a single-crystal [111]-cut $\left(2.1\times 2.3\times 0.6\right)\text{mm}^3$ diamond, synthesized using a high-temperature high-pressure (HPHT) method (Element 6). The diamond with an initial nitrogen concentration of $<$200\,ppm was electron-irradiated at 14\,MeV (dose: $10^{18}\,\text{cm}^{-2}$) and then annealed at 700$^{\circ}$C for three hours. The resulting NV centers are randomly oriented along all four \{111\} crystallographic axes of the diamond. \\
\indent The diamond is placed within a custom-made electromagnet [Fig.~\ref{Figure1}\,(a)], and is held in position with a segmented aluminum mount, which can be rotated around the z-axis and allows for optical access from both sides. The electromagnet consists of 220 turns of gauge 24 enameled copper wire and has an inner diameter of 13.5\,mm. For thermal management the wire is wound on a water-cooled aluminum spacer, and the center bore of its mount measures 8.5\,mm. The electromagnet produces 11.2\,mT/A; the current is provided by a computer-controlled power supply (Statron Typ 3257.1). The electromagnet can be moved with a computer-controlled 3D translation stage (Thorlabs PT3-Z8) to align the magnetic field with the respect to the diamond. Additionally, it can be rotated around the y-axis with a rotational stage. In combination with the possible diamond rotation, all degrees of freedom to place the diamond in the center of the magnet and to align the [111] NV-axis parallel to the magnetic field can be addressed and optimized.\\ 
\indent A secondary coil with four turns is wound around the diamond mount to apply small modulation of the magnetic field. The additional oscillating component $\text{B}_m$ is produced with the power amplified output (amplifier: AE Techron 7224-P) of a function generator (Tektronix AFG2021) that is also used as the local oscillator (LO) for the lock-in amplifier (LIA, SRS 830).\\
\indent The NV centers in diamond are optically spin-polarized with 220\,mW of 532\,nm light taken from a 12\,W laser (Coherent Verdi). The power is adjusted with a half-wave plate (HWP) and a polarizing beam-splitter (PBS) cube and transported to the optical table via a 10\,m long angle-polished, polarization maintaining fibre. Before the diamond, the light is sent trough an acousto-optical modulator (AOM, AA Electronics MT350-A0) to enable power modulation. Part of the laser light is split-off and measured on a photodiode (Thorlabs PDA36A). The signal is input into a feedback controller (PID, SRS SIM960) to stabilize the beam power. After the AOM, the beam is focused with a 40\,mm focal-length lens into the diamond. The red/near-infrared NV-PL is collected with a 30\,mm focal-length lens. The collimated PL is separated from the green transmission with a dichroic mirror and a band-stop filter for 532\,nm light before being focused with another 30\,mm lens onto a photodiode (Thorlabs PDA36A). The photodiode signal is sent to the LIA and demodulated at the LO frequency or measured at dc. 
\begin{figure}
  \centering
  \includegraphics[width=\columnwidth]{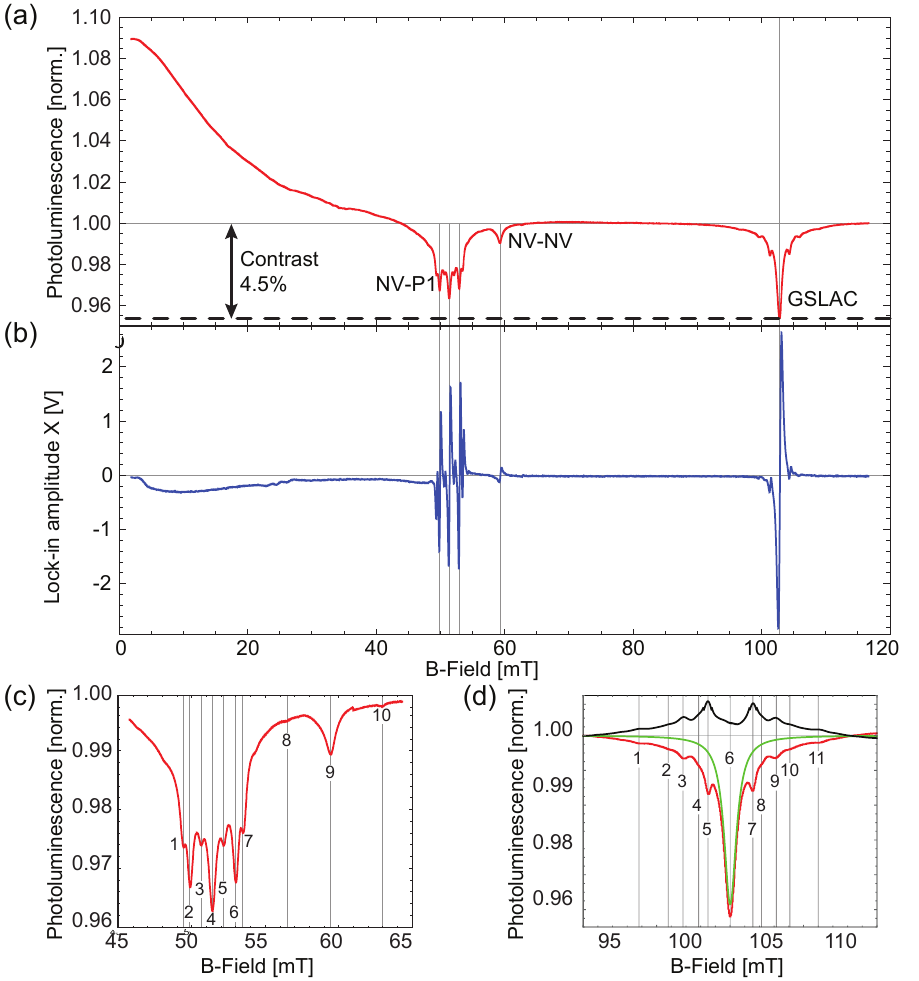}
  \caption{(a) NV-PL as a function of the applied magnetic field normalized to the PL at 80\,mT. (b) Derivative signal of (a) as given by the in-phase output (X) of the LIA. (c) Detail of the PL for fields around 51.4\,mT showing additional features. (d) Detail of the PL for fields around the GSLAC at 102.4\,mT. The data (red) weighted to ignore the side peaks are fitted with a Lorentzian function (green). The residuals of the fit are shown in black. 
}\label{Figure2}
\end{figure}
After initial alignment and calibration of the magnet, the field was scanned from 0\,mT to 120\,mT in 5\,s and the PL was monitored. Figure \ref{Figure2}\,(a) shows the PL measured with the photodiode as a function of the magnetic field. Figure \ref{Figure2}\,(b) shows the corresponding LIA signal. The modulation frequency of the field was 100\,kHz and the modulation depth $\sim$0.1\,mT, the LIA time constant 30\,$\mu$s, and 64 scans were averaged.\\ 
\indent Both plots contain several features previously discussed in the literature. The initial gradual decrease in PL is associated with a reduction in emission of the non-aligned NV centers due to spin-mixing\cite{Armstrong1}. Around 51.4\,mT, the observed features [1-7 in Fig.\,\ref{Figure2}\,(c)] correspond to cross-relaxation events between the NV center and single substitutional nitrogen (P1) centers\,\cite{Armstrong1,Hall2016,Wood2016}. 
The feature at 60\,mT [9 in Fig.\,\ref{Figure2}\,(c)] is attributed to cross relaxation with NV centers that are not aligned along the magnetic field\cite{Armstrong1}. 
At $\sim$102.4\,mT [6 in Fig.\,\ref{Figure2}\,(d)] is the GSLAC. Several additional features are visible. They can be associated with cross-relaxation events with either the nuclear spins of nearby P1 centers [2-5 \& 7-10 in Fig.\,\ref{Figure2}\,(d)]\,\cite{Armstrong1,Hall2016,Wood2016} or nuclear spins of $^{13}$C atoms [1 in Fig.\,\ref{Figure2}\,(d)].\\
\indent The angles $\alpha$ and $\beta$ between the NV-axis and the applied magnetic field need to be precisely controlled [Fig.\,\ref{Figure1}\,(a)] within $\sim$1\,mrad\,\cite{Epstein2015}. Misalignment causes a transverse field component which couples the $m_S=-1$ and the $m_S=0$ magnetic sublevels, broadens the observed GSLAC feature, and therefore leads to a reduction in magnetometric sensitivity. To optimize the GSLAC feature parameters, the angles and the position of the magnet were aligned until a minimum full width at half maximum of 1.2\,mT and an optimal contrast of 4.5\% was observed [Fig.\,2\,(b)\,\&\,(c)].\\
\indent An important characteristic of any magnetometer is the sensitivity to ac magnetic fields. For example, in eddy current sensing experiments oscillating magnetic fields need to be detected. To document the capacity of the MW-free magnetometer to detect these fields the background magnetic field is scanned around the GSLAC feature while a small oscillating magnetic field (B$_m\!\approx$\,0.09\,mT) is applied at a given frequency. The frequency is then stepped from 300\,Hz to 300\,kHz (limited by the bandwidth of the power amplifier). The PL is measured with a photodiode and its oscillating component is read-out with the LIA. On the slopes of the GSLAC feature the magnetometer is most sensitive to oscillating fields. An example of such scan is shown in Fig. \ref{Figure3}(a). For each frequency, the peak-to-peak response signal of the LIA amplitude output (R) is recorded and normalized with the oscillating current through the driving coil. This current is measured via its voltage drop over a 8\,$\Omega$ power resistor. Initially, the transfer function of the amplifier was measured, and during the experiment coarsely compensated for by adjusting the drive amplitude voltage. This way the current amplitude through the driving coil was effectively the same for all frequencies.\\
\indent Figure \ref{Figure3}\,(b) shows the peak lock-in amplitude as a function of the modulation frequency. After an initial drop in the response to oscillating fields, the spectrum appears basically flat between 60\,kHz and 300\,kHz. The initial drop in the response and the slight increase for higher frequencies can be attributed to mutual inductance of the driving coil and the surrounding background field magnet as well as induction in the aluminum magnet mount. Observation of the induced current in the main field coil is consistent with the frequency characteristic of this feature.\\%
\begin{figure}
  \centering
  \includegraphics[width=\columnwidth]{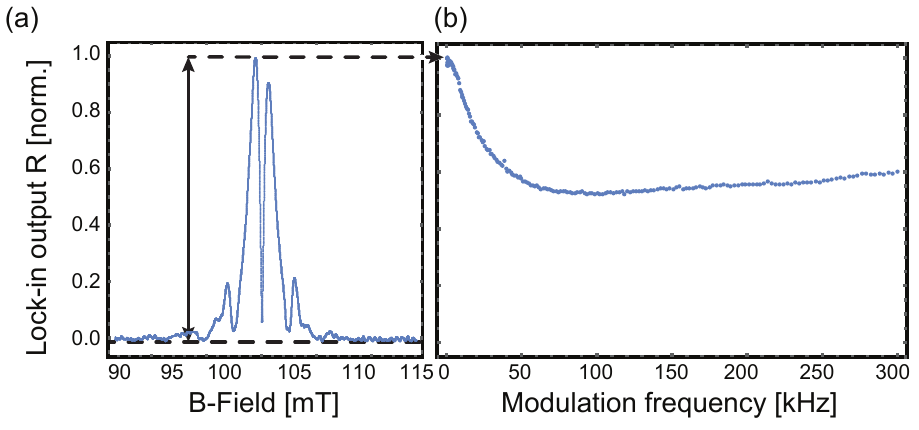}
\caption{(a) Example of the LIA amplitude output (R) for a modulation frequency of 1\,kHz. For noise reduction a moving average was applied to the data. (b) Measurement of the frequency response of the magnetometer from 300\,Hz to 300\,kHz.}\label{Figure3}
\end{figure}%
\indent Around the GSLAC feature, the derivative fluorescence signal as detected in the properly phased LIA X output depends linearly on the magnetic field and can therefore be used for precise magnetic field measurements. The calibration signal is shown in Fig.\,\ref{Figure4}\,(a); the modulation frequency (100\,kHz), modulation depth ($\sim$0.1\,mT), alignment and laser power were optimized to maximize the slope, and therefore, the sensitivity of the magnetometer. The data near the zero-crossing are then fitted with a straight line to translate the LIA output signal amplitude into magnetic field, and then the background magnetic field is set to the center of the GSLAC feature (102.4\,mT) where the magnetometer is maximally sensitive to external magnetic fields. Figure \ref{Figure4}\,(b) shows a time trace of the magnetometer response to a square-wave-modulated magnetic field of $\sim$45\,$\mu$T peak-to-peak amplitude applied with an additional external coil. The standard deviation of the data for a single step level is 1.8\,$\mu$T ($\sim$80\,msec, 4800 samples), so that the steps in the magnetic field can be observed with a signal-to-noise ratio of 25. For noise measurements, the LIA voltage output is recorded for 1\,s and translated into magnetic field variations [using the calibration performed in Fig.\,\ref{Figure4}\,(a)]. A fast-Fourier transform is performed to extract information of the magnetic field noise. The data are displayed in Fig.\,\ref{Figure4}(c) with the green pump-power stabilization (red) and without it (green). For comparison, similar data are collected at a magnetic field of around 80\,mT (blue). At this field, the setup is insensitive to magnetic field variations and the data can be used to understand the technical noise level of the magnetometer. The noise floor is flat and around 6\,nT/$\sqrt{\text{Hz}}$. The electronic noise floor without green pump light and therefore without PL (black) is about 0.25\,nT/$\sqrt{\text{Hz}}$.\\
\indent Fundamentally, the magnetometer is limited by the shot-noise of the collected PL. For the given setup, the photon shot-noise limit is calculated to be 0.43\,nT/$\sqrt{\text{Hz}}$\,\cite{AcostaPRL2013}. However, this limit could be reduced by orders of magnitude by maximizing the amount of emitted and collected PL, possible by saturating the NV-PL and increasing the numerical aperture of the collection optics.
The 1/f magnetic field noise in Fig.\,\ref{Figure4}\,(b) is attributed to the power supply that provides the main magnetic field. In an actual device however, scanning of the magnetic field would not be necessary, so that the 102.4\,mT background field could be provided by a small permanent magnet with less noise. The frequency spikes at the line voltage frequency and its higher harmonics are also attributed to the power supply. They are also the dominating noise component in Fig. \ref{Figure4}\,(b). The roll-off for frequencies above 3.5\,kHz is a result of filtering by the LIA. The time-constant for the measurements in Fig. \ref{Figure4}\,(c) was 300\,$\mu$s and the filter slope 24\,dB/octave.\\
\begin{figure}
  \centering
  \includegraphics[width=\columnwidth]{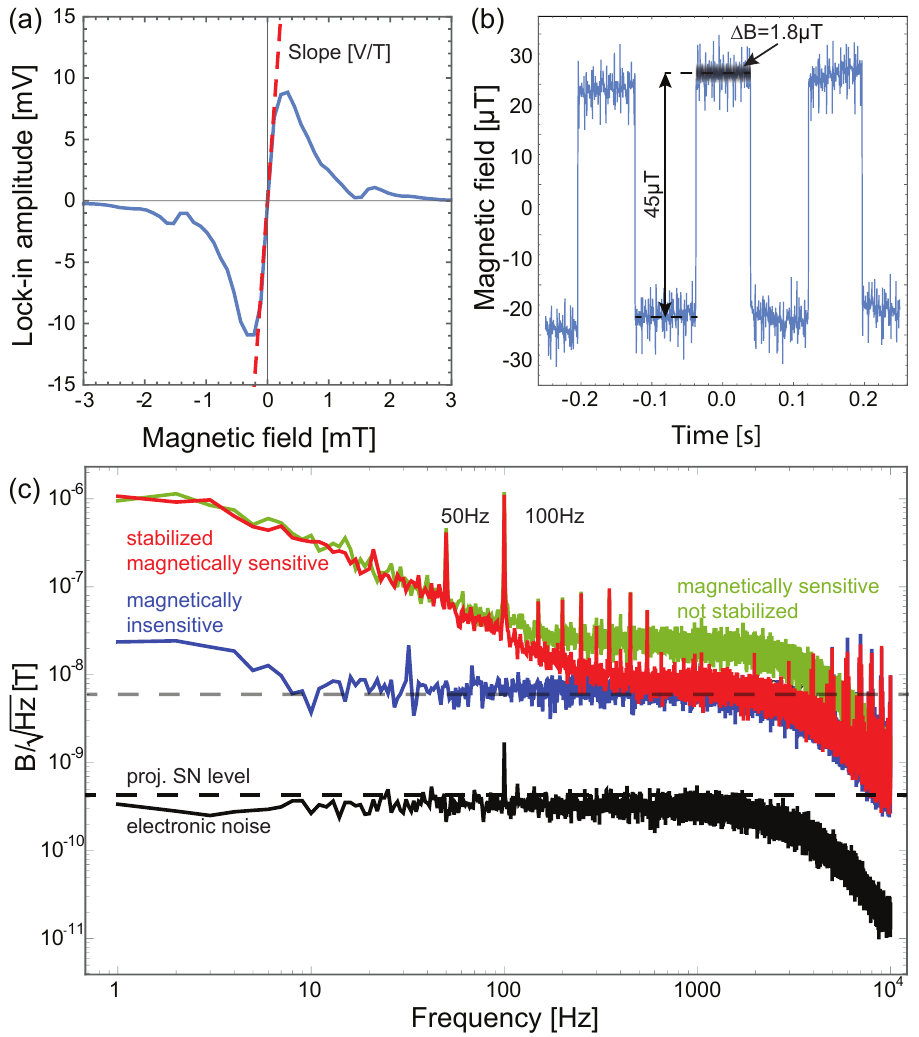}
  \caption{Magnetometer noise characterization. (a) Detail of the GSLAC feature around 102.4\,mT, fitted linearly. (b) Response of the magnetometer to a 160\,ms-period square wave magnetic field of peak-to-peak amplitude $\sim$45\,$\mu$T. The standard deviation of the magnetic field data acquired with a sampling rate of 60\,kS/s for single 80\,msec step is 1.8\,$\mu$T and indicated in gray. (c) Noise of the magnetometer: magnetically sensitive and pump intensity stabilized (red), magnetically sensitive and pump not stabilized (green), insensitive to magnetic fields with pump stabilized (blue) and electronic noise without the pump light (black).}\label{Figure4}
\end{figure}
In conclusion, we demonstrated a MW-free, NV-center based magnetometer with a 6\,nT/$\sqrt{\text{Hz}}$ noise floor and a bandwidth exceeding 300\,kHz. This device can be useful in applications where microwave spectroscopy cannot be performed on the NV centers. This is the case where the diamond-based sensor is placed in proximity to conductive objects, and as such, is of particular relevance for spatially resolved conductivity measurements in the context of magnetic induction tomography. The ability of the present technique to detect nuclear spins (seen as side-features near the GSLAC peak) with high signal-to-noise ratio indicates a possibility of applications in sensing spins external to the NV centers. If a layer of shallow-implanted NV-centers is used, spins external to diamond can be probed.\\
\indent Future investigations will involve a thorough study of the lineshape and width of the signal near the GSLAC, as well as of the additional features around it, with the aim of understanding the fundamental limitations of our sensing protocol. In addition, combination of the presented MW-free magnetometer with an absorption-based protocol will allow for magnetic field sensing with a sensitivity exceeding the PL shot noise limit\cite{NVCavity1,Cavity2}.\\
\begin{acknowledgements} 
We acknowledge support by the DFG through the DIP program (FO 703/2-1). HZ is a recipient of a fellowship through GRK Symmetry Breaking (DFG/GRK 1581). NL acknowledges support from a Marie Curie International Incoming Fellowship within the 7th European Community Framework Programme. LB is supported by a Marie Curie Individual Fellowship within the second Horizon 2020 Work Programme. DB and AJ acknowledge support from the AFOSR/DARPA QuASAR program. V.M.A. acknowledges support from NSF Grant No. IIP-1549836. We thank P.\,R.~Nelson, J.\,W.~Blanchard, and D. Twitchen for useful discussions.
\end{acknowledgements}
\bibliography{literature}
\end{document}